\RequirePackage{fix-cm}
\documentclass[twocolumn]{svjour3}          
\smartqed  
\usepackage{graphicx}
\usepackage{bbding}
\usepackage{array}
\usepackage{url}
\begin{document}

\title{Software-Defined and Virtualized Future Mobile and Wireless Networks: A Survey
\thanks{This work is supported by National Basic Research Program of China (973 Program Grant No.~2013CB329105), National Natural Science Foundation of China (Grants No.~61301080 and No.~61171065),
Chinese National Major Scientific and Technological Specialized Project (No.~2013ZX03002001), China¡¯s Next Generation Internet (No.~CNGI-12-03-007), and ZTE Corporation.}
}

\author{Mao Yang \and
        Yong Li \and
        Depeng Jin \and
        Lieguang Zeng \and
        Xin Wu \and
        Athanasios V. Vasilakos \and
}


\institute{M. Yang\at
              School of Electronics and Information, Northwestern Polytechnical University, Xi'an 710072, P. R. China\\                      
              \email{yangmao210@gmail.com}
           \and
           Y. Li\Envelope $\cdot$ D. Jin $\cdot$ L. Zeng \at
              Department of Electronic Engineering, Tsinghua University, Beijing 100084, P. R. China
                      \and
           Y. Li\Envelope  \at
              \email{liyong07@tsinghua.edu.cn}
           \and
           D. Jin, L. Zeng \at
              \email{\{jindp, zenglg\}@mail.tsinghua.edu.cn}
           \and
           Xin Wu \at
              Big Switch, USA\\
              \email{xin.wu@bigswitch.com}
           \and
           A. V. Vasilakos \at
              Department of Computer and Telecommunications Engineering,University of Western Macedonia, Greece\\
              Electrical and Computer Engineering, National Technical University of Athens (NTUA), Greece\\
              \email{vasilako@ath.forthnet.gr}
}

\date{Received: date / Accepted: date}

\maketitle

\begin{abstract}
With the proliferation of mobile demands and increasingly multifarious services and applications, mobile Internet has been an irreversible trend. Unfortunately, the current mobile and wireless network (MWN) faces a series of pressing challenges caused by the inherent design. In this paper, we extend two latest and promising innovations of Internet, software-defined networking and network virtualization, to mobile and wireless scenarios. We first describe the challenges and expectations of MWN, and analyze the opportunities provided by the software-defined wireless network (SDWN) and wireless network virtualization (WNV). Then, this paper focuses on SDWN and WNV by presenting the main ideas, advantages, ongoing researches and key technologies, and open issues respectively. Moreover, we interpret that these two technologies highly complement each other, and further investigate efficient joint design between them. This paper confirms that SDWN and WNV may efficiently address the crucial challenges of MWN and significantly benefit the future mobile and wireless network.

\end{abstract}

\section{Introduction} \label{sec:introduction}
Over the past few decades, mobile and wireless communications as well as Internet are the most profound and important technology in information technologies that rapidly grow and continuously change human life. It seems that these ``twins'' kept developing in a parallel and independent ways. Specifically, the information theory, \emph{i.e. Shannon Theory}, is considered as one of the most fundamental theory in communications\cite{Cover}, while Internet evolves in practice from laboratory to globalization \cite{History_Internet_Magazine,History_Internet_Sigcomm}. Nevertheless, recent years, the demands of mobile data keep proliferating, meanwhile, the smart phone and mobile applications become increasingly powerful and multifarious. These lead to a rare opportunity to combine both mobile communications and Internet, \emph{i.e.} mobile Internet. Mobile Internet enables us to enjoy diverse Internet services through mobile and wireless accessing such as online video, file sharing, web browsing, voice service, and \emph{etc}. Consequently, mobile and wireless network (MWN) is changing from voice-centric to multimedia-oriented \cite{Multimedia_Centric_3G,Multimedia_Centric_Energy}.

Unfortunately, traditional mobile and wireless network can hardly keep up with this trend and currently face several intractable challenges: a) the contradiction between the increasing scarce spectrum and the poor resource utilization \cite{Spectrum_Conflict} aggravates gradually. b) There are tremendous heterogeneous wireless networks, but it is difficult to achieve efficient interworking between them to meet end users' diversified expectations while ensuring the coexistence of diverse heterogeneous networks\cite{Heterogeneous}. c) Tightly coupling with specific hardware and lack of flexible control interfaces, the current mobile network can hardly provide a fast track for technological innovations. d) Mobile services and applications conspicuously become multifarious, and they require different types of network characteristics\cite{Mobile_QoS_QoE}. Unfortunately, traditional MWN only support these multifarious services with the same network characteristic. As a result, it naturally deteriorates the quality of service (QoS) and quality of experience (QoE) of end users. e) Carriers sink into a predicament that the network costs keep growing while the revenues remain stagnate, which is inconsistent with the service and application proliferation\cite{Mobile_Revenue}\cite{Mobile_Cost}. These challenges directly affect all the participants including academia, carriers and end users. Consequently, they impede the evolution of future MWN.

Most of these pressing challenges are deeply rooted in the inherent design of the current mobile communication and Internet. For example, vertical-constructing method adopted by mobile communication hampers the convergence of heterogeneous networks; the distributed and independent operating model results in low resource utilization; the tightly coupling between hardware and wireless protocols makes MWNs difficult to evolve and deploy new technologies and functions; mobility becomes increasingly complicated and hard to handle because IP address based Internet is oriented towards the fixed network from its birth. Consequently, simply evolving the current mobile networks can hardly meet such great expectations without fundamental architectural changes. On the other hand, although clean-slate design is important for enabling the networking field to mature into a true discipline\cite{Cleanstate_vs_evolution}, it is entirely incompatible with the current technologies and may not be acceptable due to the incalculable costs. Therefore, a revolutionary architecture is required to enable the MWN to possess the capability to address these pressing challenges and to achieve smoothly evolution.

Software-defined network (SDN)\cite{SDN_Jennifer,CM_SDN_Management}, an innovative paradigm, is one of the latest and hottest topics in networks. SDN advocates decoupling the control plane and data plane of networks, which dramatically simplifies network control and enables innovation and evolution by abstracting the control functions of the network into a logically centralized control plane. Network devices become simple packet forwarding and processing devices (the data plane) that can be programmed via an open interface\cite{SDN_History}, such as OpenFlow\cite{Openflow_Survey,SDN_Openflow}. Importantly, in recent years, academia and industry lead one new trend that extending SDN to mobile and wireless networks, i.e. software-defined wireless network (SDWN). Orienting towards the expectations and challenges of MWN, SDWN benefits several aspects of MWN, for example wireless resource optimizing, convergence of heterogeneous networks, fine-grained controllability, and efficient programmability for network innovation and smooth evolution. However, key technologies in software defined wireless networks (SDWN) are still need to be addressed, such as the end-to-end architecture, physical layer software defining, control strategies, and etc.

Network virtualization\cite{Network_Virtualization,Network_Virtualization_2} is currently attracting ubiquitous attention from both academia and industry. It enables multiple concurrent virtual networks to run on shared substrate resources. Network virtualization separates the traditional Internet service providers (ISPs) into two new roles: infrastructure network providers (InPs) and service providers (SPs)\cite{Network_Virtualization_Inp_SP} and, consequently, SPs require virtual networks to offer various services by leasing resources from from InPs. Recent years, some researchers focus on wireless network virtualization (WNV). It significantly improves the resource utilization, benefits the network innovations, offers the customized services, and enhances the QoS and QoE.

With the rapid growth of mobile demands and the ever-increasing diversity of services and applications, mobile Internet has been an irresistible trend. On the other hand, software-defined network and network virtualization emerge and offer a great opportunity to integrate communications and Internet. Although software-defined wireless network and wireless network virtualization are considered as different technologies, they highly complement each other. It is quite necessary to combine SDWN and WNV to efficiently address the challenges and provide opportunities for future mobile and wireless network, for example the Network Function Virtualization (NFV) group\cite{NFV}\cite{NFV_SDN}. This paper focuses on the software-defined and virtualized future mobile and wireless networks. Firstly, We analyze the crucial challenges of current MWN and the possible opportunities provided by SDWN and WNV. Then, we respectively describe the main idea, ongoing researches and key technologies, and challenges and open issues of SDWN and WNV in detail. We further study the effective combination of them. Therefore, we highlight that this paper investigates two potential technologies in mobile and wireless network, which provides the future networks with possible solutions and exploits promising directions for the future research.

The remainder of this article is organized as follows. We first offer the challenges of current MWN and the opportunities provided by SDWN and WNV in Sec. \ref{sec:Challenges}. Next, Sec. \ref{sec:SDWN} and Sec. \ref{sec:Virtualization} present the technical details of SDWN and WNV respectively. Then, the possible effective cooperation of them is studied in Sec. \ref{sec:Combination}. In Sec. \ref{sec:Conclusion} we conclude this paper.

\section{Challenges and opportunities for future mobile and wireless network} \label{sec:Challenges}
The intractable plights that mobile and wireless network facing are mostly caused by the inherent design and the proliferating demands. Simply evolving the traditional mobile communication system can hardly meet the future demands. This section surveys the challenges mobile and wireless network has to face, and investigate the opportunities that SDWN and WNV bring.

\noindent \textbf{Challenge 1: Diverse heterogeneous networks \emph{v.s.} vertical constructing and operating}. There are diverse heterogeneous wireless networks, such as LTE, Wimax, UMTS, and WLAN. These heterogeneous networks will coexist for a long time\cite{Ericsson}, and each of them has specific characteristics suitable for different services. For example, 3G offers high-quality voice services and WLAN provides better high speed video services. However, heterogeneous networks can hardly interconnect due to the vertical constructing and operating strategy rooted in the inherent design, which makes operators incapable of achieving complementary advantages and efficiently optimizing from the global perspective.

\textbf{Opportunities:} The simplified network devices in SDWN dynamically report the runtime network status of heterogeneous networks. Then the controller easily possesses the global network views and, consequently, schedules the forwarding rules and data processing strategies. Therefore, this important separation enables the interconnections among heterogeneous networks. On the other hand, the control plane may dynamically optimize multiple heterogeneous networks such as resource allocation based on the realtime network requirements, effective offloading\cite{Offloading_SDN}, and other efficient network cooperative strategies.

\noindent \textbf{Challenge 2: Capability crisis \emph{v.s.} low resource utilization}. Several big carriers, including AT\&T, Verizon and Sprint, argue that in the next few years they may not have enough spectrum to meet the mobile demands\cite{Spectrum_Crisis}; meanwhile, it seems that mobile and wireless network encounters ``capability crisis'' that the network capacity will not keep up with the mobile demands proliferating\cite{Capacity_vs_Demands}. Many scientists and engineers are convinced that the main reason for spectrum crisis is that wireless resources are not fully utilized\cite{Spectrum_Crisis}. As a result of the difficulty in achieving the convergence of heterogeneous networks, many network devices are not fully utilized and plenty of wireless resources are wasted.

\textbf{Opportunities:} WNV enables operators to provide services by requiring, managing and operating virtual networks sharing the same substrate physical networks. As a result of this sharing, it significantly improves the resource and device utilization. In the meantime, SDWN may dynamically optimize and adjust the resources utilization according to the global network characteristic, service requirements variation, and the real time network statuses from the global perspective.

\noindent \textbf{Challenge 3: Innovation expectations \emph{v.s.} network ossification}. The increasingly rapid development and intractable challenges compel the mobile and wireless network to continuously innovate. Unfortunately, as mobile network technologies are usually solidified in the hardware, i.e. network ossification, it is impossible for researchers to verify and further improve their network innovations in the real environment, and it conspicuously impedes the technologies deployment and network evolving.

\textbf{Opportunities:} SDWN efficiently addresses the network ossification and makes mobile and wireless network programmable, in term of both the control plane and data plane. Consequently, innovative technologies can be flexibly deployed and verified by programming though open APIs. Meanwhile, WNV may construct independent virtual networks to meet the expectations of innovations without interfering other inservice virtual networks. Furthermore, SDWN and WNV makes new technologies and ideas easy and fast to deploy, which naturally benefits the network evolving smoothly.

\noindent \textbf{Challenge 4: Service and application proliferating \emph{v.s.} poor QoS and QoE}. Wireless services and applications proliferate significantly and become increasingly multifarious. Different kinds of services always require very different types of network characteristics. However, current MWN only supports these wide-ranging services with the same network characteristic. Naturally, it deteriorates the quality of service (QoS) and quality of experience (QoE) of end users.

\textbf{Opportunities:} WNV is able to provide various services with several virtual networks with appropriate characteristics guarantees. In this case, it is easy for us to deploy customized services. At the same time, the centralized control plane in SDWN may dynamically adjust the network configurations in terms of the realtime QoS and QoE levels since QoE has received increasing interests from both academia and users.

\noindent \textbf{Challenge 5: Rapid growth of traffic and subscribers \emph{v.s.} increasing cost and stagnant revenues}. Although the growth of mobile data are forecasted to double annually in the next few years, carriers sink into a predicament that the network costs keep growing but the revenues is likely to remain stagnate\cite{Mobile_Revenue}. The operators urgently call for a new operating model in order to continuously increase the revenues and reduce the costs.

\textbf{Opportunities:} As mentioned above, SDWN and WNV improve the resource utilization and save the costs by globally optimizing and resource sharing, such as energy and device utilizing cost. More importantly, they introduce new operating models for the future networks. For example, WNV decouples the resource provision and service provision, and introduces two new operating entities InPs and SPs; SDWN enables the operators to provide open interfaces to the service providers.

Therefore, software-defined wireless networks and wireless network virtualization are promising ways to address the crucial challenges of the future mobile and wireless networks. In the next sections, we will describe SDWN and WNV in details.

\section{Software-defined wireless network (SDWN)} \label{sec:SDWN}
\subsection{Software-defined network}
SDN separates the data plan and control plan of the network and introduces a logically centralized control plane, referred to as SDN controller, to abstract the control functions of network\cite{SDN_Jennifer,CM_SDN_Management}. As a result, network devices are simplified to a great grade, and their packet forwarding and data processing functions can be programmed via an open interface. Thus, SDN considerably simplifies the network devices and makes networks more controllable and flexible.

Plenty of network operators, service providers, and vendors founded and joined Open Network Foundation (ONF)\cite{ONF}, an industrial-driven organization, and Open Networking Research Center (ONRC) \cite{ONRC}, an research-oriented organization. OpenFlow\cite{Openflow_Survey}\cite{SDN_Openflow}\cite{OpenFlow_Specification} is the most recognized realization of SDN. The OpenFlow protocol is sustainably released by ONF. Many vendors, such as HP, NEC, IBM, design and develop the available commercial OpenFlow switch, and several kinds OpenFlow controllers, e.g. NOX\cite{NOX} and floodlight\cite{Floodlight}.

One important but underlying reason for the wide attentions for SDN lies in that SDN is just a paradigm, rather than an ossified architecture. In other word, it is reasonable for researchers to study the technologies for networking by adopting the core idea of SDN, not being limited in one fixed structure. As a result, SDN has been applied or extended to various scenarios, such as data center networks\cite{SDN_DCN_Google}\cite{SDN_DCN_Microsoft}, network security\cite{SDN_Security_1}\cite{SDN_Security_2}, optical networks\cite{SDN_Optical_1}\cite{SDN_Optical_2}\cite{SDN_Optical_3}, and naturally, mobile and wireless scenarios studied in this paper.

\subsection{Software-defined wireless network (SDWN)}
\begin{figure}
\centering
  \includegraphics[width=0.50\textwidth]{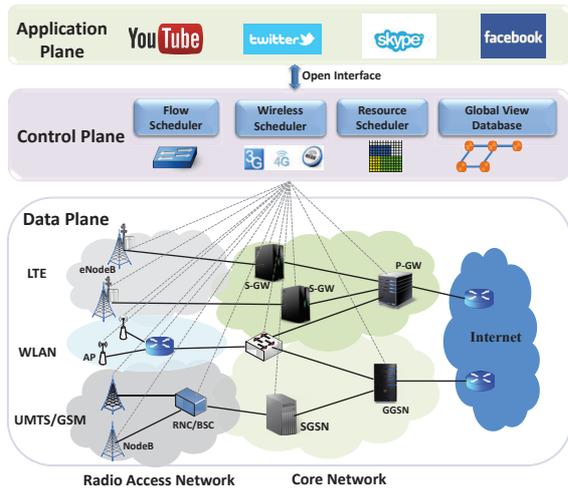}
\caption{Schematic architecture of software-defined wireless network}
\label{fig:architecture}
\end{figure}

SDN brings various advantages for the Internet. As the mobile demands keep proliferating, mobile and wireless networks eventually become the bottleneck of MWN, thus it is necessary to extend the SDN paradigm to the MWN. Over recent years, software-defined wireless network (SDWN) has become an emerging and significant research branch for SDN, and captured increasingly attentions. Orienting towards the characteristics of mobile and wireless network, SDWN aims to study the network architecture and a series of relative crucial technologies for the future mobile and wireless network. Fig. \ref{fig:architecture} depicts the conceptual schematic architecture of SDWN. Mobile and wireless network is composed of two dimensions, both ``north-south'' and ``east-west''. For ``north-south'', MWN consists of radio access network and core network; while for the ``east-west'' dimension, there are multiple heterogeneous networks. Consequently, deploying software-defining paradigm in wireless scenario is more complicated than that in the wired networking.

Consistent with the core idea of SDN, SDWN separates the control plane and the data plane. The network devices, including radio accessing devices, forwarding devices, and etc., have been simplified and behave according to the rules scheduled by the logically centralized control plane. Possessing the global information dynamically reported by the network devices, the control plane allocates the resources, schedules the devices behaviors, and configures the wireless parameters. SWDN benefits all the the network entities---network operators, service providers, and end users.

1) Network operators. SDWN enables the control plane to obtain, update, even predict the global information, such as the users' attributes, network requirements variation, and the realtime global network status. Correspondingly, the control plane is able to optimally schedule and/or adjust the resource allocation, forwarding strategies, wireless configuration, and etc., by sending rules and altering the behaviors of network devices. Therefore, SDWN makes networks more controllable and programmable for the network operators.

2) Service providers. SDWN abstracts the network functions and provides multiple open APIs. Accordingly, service providers are able to offer more multifarious and customized services by leveraging modular abstractions and open APIs. Meanwhile, QoS can be guaranteed by efficiently programming.

3) End users. Firstly, SDWN directs the flows along the most appropriate paths from or to the end users in view of the users' attributes, the application types, and the network status. Moreover, it guides the end user to select the best accessing network, even enable the end users to be served by multiple networks.

More importantly, SWDN efficiently \textbf{addresses the pressing challenges} of MWN mentioned in Sec \ref{sec:Challenges}. First, decoupling the control plane and data plane conspicuously simplifies network devices and makes them programmable, which promotes the interconnection among heterogeneous networks by dynamically scheduling the forwarding rules. On the other hand, SDWN optimizes the resource allocation among networks based on the requirements and network status, and helps end users to access to the best network or simultaneously be served by multiple heterogeneous networks. Therefore, SDWN benefits the \emph{convergence of heterogeneous wireless networks}. Second, acquiring the realtime global information makes SDWN easy to \emph{enhance the resource utilization} and \emph{reduce the costs} through reasonable operating rules and resource allocations. Third, effective programmability and diversified open APIs enable network to efficiently deploy the new technologies and smoothly evolve by programming, which meets the \emph{network innovations} expectations. Fourth, as SDWN is beneficial to network operators, service providers, and end users, it naturally \emph{increases the revenues} of each entity. For example, QoS of service providers, the resource utilization of network operators, QoE of end users, and of course, economic benefits of them.

\subsection{Ongoing research on SDWN}
Currently, SDWN is still in its infancy, and most related works mainly focus on the SDWN architecture.

\begin{table*}
\newcommand{\tabincell}[2]{\begin{tabular}{@{}#1@{}}#2\end{tabular}}
\caption{Representative summary of studies on SDWN}
\label{tab:SDWN}
\begin{tabular}{r@{}lr@{}lr@{}lr@{}lr@{}l} \hline
\multicolumn{2}{c}{} &&year &&network type && network position && summary\\ \hline
\multicolumn{2}{c}{OpenRoad\cite{OpenRoad_1}} && 2009 && Wifi     && OpenFlow switch && \tabincell{l}{OpenFlow based open-source platform for moving freely\\ between any wireless infrastructure}\\
\multicolumn{2}{c}{Odin\cite{Odin}}           && 2012 && WLAN     && OpenFlow switch && \tabincell{l}{Prototype software defined wireless network framework for\\ enterprise WLANs by introduces progr-\\ammability}\\
\multicolumn{2}{c}{OpenRadio\cite{OpenRadio}} && 2012 && General  && Access network  && \tabincell{l}{A programmable wireless data plane for the continuous\\ wireless network evolving}  \\
\multicolumn{2}{c}{OpenRF\cite{OpenRF}}       && 2013 && WLAN     && Access network  && \tabincell{l}{A software-defined cross-layer architecture for managing\\ MIMO signal processing with commodity\\ Wi-Fi cards} \\
\multicolumn{2}{c}{SoftRAN\cite{SoftRan}}     && 2013 && LTE      && Access network  && \tabincell{l}{A software-defined centralized control plane of the radio\\ access network by introducing a virtual\\ big-base station}  \\
\multicolumn{2}{c}{OpenRAN\cite{OpenRAN}}     && 2013 && Heterogeneous && Access network && \tabincell{l}{A cloud computing based software-defined radio access\\ network architecture for heterogeneous wireless networks} \\
\multicolumn{2}{c}{CellSDN\cite{CellSDN}}     & &2012 && LTE      && Access + core   && \tabincell{l}{Abstract the control functions from both accessing and\\ forwarding devices to meet the needs for fast and frequent\\ updates}  \\
\multicolumn{2}{c}{SoftCell\cite{SoftCell}}   && 2013 && LTE      && Core network && \tabincell{l}{Software-defined cellular core network by aggregating traffic\\ from multi-dimensions and simplifying the P-GW}  \\
\multicolumn{2}{c}{MobileFlow\cite{MobiFlow}} && 2013 && LTE      && Carrier network && \tabincell{l}{Carrier-grade flow-based forwarding and a rich environment\\ for innovation at the core of the mobile network}  \\
\multicolumn{2}{c}{SoftCOM\cite{SoftCOM_1}}   && 2013 && Heterogeneous && Access + core && \tabincell{l}{Cloud-based systematic future network architecture that\\ leverages SDN and enhances telecom operators' competitiveness\\ by transforming network, operations and business}  \\
\hline
\end{tabular}
\end{table*}

OpenRoad\cite{OpenRoad_1}\cite{OpenRoad_2} is the first work about SWDN. OpenRoad envisages a world in which users can move freely between
any wireless infrastructure by separating the network service from the underlying physical infrastructure. They propose an OpenFlow based open and backward compatible wireless network infrastructure, and deploy it in college campuses\cite{OpenRoad_3}. Network devices are controlled by NOX. They also introduce virtualization by FlowVisor\cite{FlowVisior_1}\cite{FlowVisior_2}. Finally, they verify the handover performance among Wifi and WiMax. However, OpenRoad manly orients towards WiFi and provides no special support for cellular networks. Odin\cite{Odin} is a prototype software defined wireless network framework for enterprise WLANs by introduces programmability. Odin builds on a light virtual access point (AP) abstraction, virtualizing association state and separate them from the physical AP. The decision module in Odin is an application on top of the OpenFlow controller. Odin benefits several applications such as seamless mobility, load balancing and hidden terminal mitigation.

The fast-varying and vulnerable mobile and wireless environment makes the PHY and MAC layers technologies in mobile networks much more complicated than in the wired scenarios. In recent years, some studies concentrate on investigating lower layer wireless network technologies in the SDWN. OpenRadio\cite{OpenRadio} proposes a programmable wireless data plane, implemented in multi-core hardware platform, for the continuous wireless network evolving. In the current wireless network, the network operators have no choice but to replace the devices because wireless stacks are usually solidified in the hardware. Differently, OpenRadio firstly introduces a software abstraction layer and offers a modular and declarative APIs. OpenRadio decomposes any specific wireless protocols into processing plane and decision plane components with these APIs. Concretely, the decision plane schedule the ``rules'' by encapsulating all the decision logic functionality. The processing plane encapsulates a group of signal processing algorithms used in a PHY processing chain such as FFT, 64-QAM, convolutional encoding, Viterbi decoding, and etc. Therefore, OpenRadio implements the PHY and MAC layers of different wireless protocol stacks such as WiFi, LTE and WiMax by flexibly programming. OpenRF\cite{OpenRF} proposes a software-defined cross-layer architecture for managing MIMO signal processing in today's networks with commodity Wi-Fi cards. OpenRF adopts the SDN idea and enables access points to control MIMO signal processing at physical layer, such as interference nulling, coherent beamforming and interference alignment. After converting high-level quality of service requirements of downlink flows to low-level physical-layer techniques, OpenRF controller makes cross-layer decision to guarantee the transmission rate and control interference across access points.

Software-defined radio access network is also an attractive direction of SDWN since radio access network (RAN) provides ubiquitous wireless connectivity to mobile end users and integrates the mobile and wireless features. SoftRAN\cite{SoftRan} is a software-defined centralized control plane of the radio access network by introducing a virtual big-base station. The virtual big-base station consists of a central controller and radio elements (individual physical base stations). To achieve the tradeoff between the optimal centralized control and the sensitive inherent delay, SoftRAN redesigns the control plane functionalities cooperatively between the the controller and the radio elements. Specifically, the centralized controller handles the cross radio elements decision, while the individual radio elements deal with
the frequently varying parameters. SoftRAN mostly focuses on the one-tier situation, consisting of microcells, which is not suitable for the scenarios of heterogeneous networks. OpenRAN\cite{OpenRAN} proposes a software defined radio access network from another perspective by introducing cloud computing inspired from C-RAN\cite{C_RAN}. OpenRAN consists of wireless spectrum resource pool, cloud computing resource pool and a SDN controller. The wireless spectrum resource pool covers multiple heterogeneous wireless networks. The cloud computing resource pool implements the baseband processing of these heterogeneous networks. According to the dynamic network requirements, SDN controller establishes the virtual base station in the wireless spectrum resource pool, and corresponding virtual baseband processing unit in the cloud computing pool by installing appropriate PHY and MAC layer protocols.

As core network in mobile and wireless centralizes almost all data-plane functionalities and traffic, CellSDN\cite{CellSDN} and SoftCell\cite{SoftCell} aim at studying the software-defined core network in the LTE network. CellSDN covers both access and core network, and deploys a network operating system to abstract the control functions from both accessing and forwarding devices. To meet the demands for fast and frequent updates, it introduces a local agent to make the realtime decision. SoftCell is the successive research of CellSDN, and mainly focuses on the challenges of core network. SoftCell aggregates traffic from three dimensions---the service policy, the base station location, and the UEs. SoftCell simplifies the packet gateway (P-GW) through two ways: introducing the software-defined access switches to implement the packet classification, and configuring optimal forwarding paths across various specific middleboxes. MobileFlow\cite{MobiFlow} studies the carrier network and proposes blueprint for flow-based forwarding and a rich environment for innovation at the core of the mobile network.

SWDN also attracts attention from equipment vendors. SoftCOM\cite{SoftCOM_1}\cite{SoftCOM_2}, proposed by Huawei, is a cloud-based systematic future network architecture that leverages SDN and enhances telecom operators' competitiveness in transforming network, operations and business.

Moreover, There are some studies referring to SWDN for mesh or ad-hoc networks\cite{SDN_Adhoc_1}\cite{SDN_Adhoc_2}. A representative summary of SWDN studies is shown in Table \ref{tab:SDWN}.

\subsection{Challenges and open issues}
\subsubsection{Supporting heterogeneous networks}
One specific wireless standard dominating the future mobile and wireless network will be nearly impossible for a long time, which requires the SDWN to efficiently support these diverse standards by guaranteeing their characteristics, ensuring fast deployment, and enabling smoothly evolution.

\subsubsection{Cross layer software-defining}
Since the PHY and MAC layers technologies in MWN are much more complicated than in the wired scenarios, it is indispensable for SDWN to achieve cross layer software-defining. Reasonably decomposing and abstracting the control functions from the network layer down to the physical layer is one fundamental challenges for SDWN. It also requires SDWN to possess effective programmability.

\subsubsection{Control strategy design}
Multifarious and ever-changing mobile and wireless environment naturally leads to complex control. For example, the application diversity and fast mobility result in fast-varying network requirements and status and, correspondingly, control difficulties. Therefore, SDWN needs to find a flexible and effective control strategies.

\subsubsection{Open Interfaces}
SDWN abstracts the control functions and provides open APIs for service providers. As the services and applications keep proliferating and diversified, it is obviously one pressing challenge that SDWN should firstly analyze the service characteristics and the network capabilities and, accordingly, offer appropriate, concise, and flexible open interfaces for service providers.

\subsubsection{Real-time and mobility property}
Compared with wired scenario, mobile networks always require lower latency because of the traditional mobile services, e.g. voice. On the other hand, fast-varying and vulnerable wireless properties further impede the QoS and QoE of real-time property. These obviously challenge SDWN due to the logically central controlling. Furthermore, mobility is one of the most fundamental and important features in mobile networks, especially for the future dense deployed wireless networks. Thus, to efficiently support the fast mobility is also an open issue to be addressed.

\subsubsection{Scalability}
There are some studies focusing on the scalability of wired SDN\cite{SDN_Scalability}\cite{SDN_Scalability2}, but it is difficult in the wireless scenario. As the size of a network enlarges, more packets are sent to the controller. There is no doubt that the controller can hardly handle all these incoming requests\cite{SDN_Scalability}. Simply improving the performance of the sole centralized controller, without designing from the architecture level, can impossibly adapt to the wide-ranging and increasing dense network scale. Therefore, SDWN might consist of multiple controllers physically distributed in the system. These controllers will not conflict with the ``logically centralized'' principle through communicating and cooperating with each other efficiently.

\subsubsection{Virtual Machine (VM) Migration}
In SDN world, one of the most challenging tasks is to track VM movement, i.e., if a VM moves from one switch to another, the SDN controller has to detect this movement and reprogram the forwarding entries of that VM on every single switch. In mobile networks, this problem will be even worse because the complexity of mobile environments and the frequent handovers caused by the high speed mobility.

\section{Wireless network virtualization (WNV)} \label{sec:Virtualization}
\begin{figure}
\centering
  \includegraphics[width=0.50\textwidth]{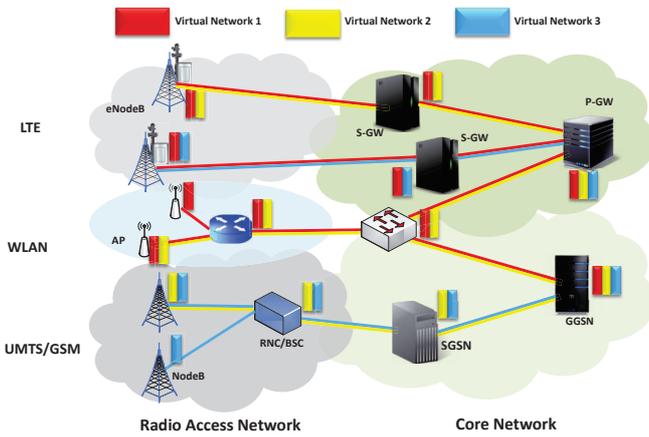}
\caption{Schematic architecture of wireless network virtualization}
\label{fig:WNV}       
\end{figure}

\subsection{Network virtualization}
Network virtualization\cite{Network_Virtualization}\cite{Network_Virtualization_2} has rapidly developed as a promising paradigm to solve the inherent problems faced by the current Internet\cite{Network_Virtualization_Inp_SP}. By enabling multiple isolated virtual networks to coexist on the shared physical substrate, network virtualization decouples the traditional ISPs into two independent entities: infrastructure providers (InPs), who manage the physical substrate, and service providers (SPs), who lease resources from InPs and offer end-to-end services by creating and operating virtual networks (VNs).

Network virtualization has been widely used in many aspects of networking, which achieves higher resources utilization, lower cost, more flexible management, and higher energy efficiency in data centers\cite{NV_DataCenter}\cite{NV_DataCenter2} and cloud computing\cite{NV_CloudComputing}\cite{NV_CloudComputing2}. Meanwhile, it is also a principle technologies of future network testbeds such as GENI\cite{GENI}, PlanetLab\cite{PlanetLab}, FIRE\cite{FIRE}, and TUNIE\cite{TUNIE}.

\subsection{Wireless network virtualization (WNV)}
Recently, network virtualization has been extended to mobile and wireless network scenario. Fig. \ref{fig:WNV} depicts the conceptual schematic architecture of wireless network virtualization (WNV). Similarly, WNV enables several concurrent virtual networks running on the shared wireless physical substrate, and introduces two entities: InPs and SPs. The InPs possess physical mobile network infrastructure and operate the physical network, while the SPs lease network resource from InPs and further create virtual networks to provide various mobile services. Generally, the network substrate includes nodes (network devices) and links. Therefore, in WNV, the physical nodes and physical links are virtualized into several virtual nodes and virtual links belonging to different virtual networks.

However, WNV is quite different from and more complex than wired network virtualization due to the unique wireless features. First, multiple heterogeneous wireless networks requires WNV to possess the ability to support several wireless protocols simultaneously, which is challenging because the key technologies and performance parameters among multiple protocols are quite different. Second, the mobile and wireless environment makes the network status fast-varying. Correspondingly, it is not easy for WNV to dynamically schedule the resources, configurations, and protocols. Third, mobile services and applications become increasingly multifarious, in addition to the randomness of user behavior, which consequently makes it difficult to timely create appropriate virtual networks for these ever-changing services. WNV offers several advantages for the mobile and wireless network.

1) Efficient resource utilization in heterogeneous networks. Multiple heterogeneous networks may independently run in different virtual networks by guaranteeing the resource and protocol and ensuring the isolation feature. Meanwhile, service providers on longer have to construct their own networks because they just need to put forward their requirements and lease resources from InPs. Therefore, WNV significantly benefits the heterogeneous network convergence, improves the resource utilization and achieves the energy efficiency.

2) Customized and high QoS/QoE service supply. Different types of services and applications may be supported in different virtual networks. For example, one virtual network carries voice services for its low latency feature, while another one deploys video stream services by providing high transmission rate. Consequently, WNV is able to provide customized services and guarantee the QoS and QoE.

3) Innovation platform. Wired and wireless network virtualization is naturally one of the most capable technologies for the network innovation testbeds and, in fact, there are already lots of platforms currently such as GENI\cite{GENI} and ORBIT\cite{ORBIT}. Researchers may deploy and verify their innovations through practical experiments.

4) New economic model. WNV decouples the physical resource provision and service provision, and brings two new entities: InPs and SPs. As a result, both of them focus on their own responsibilities. Consequently, this increases the revenue and reduces the cost for both of them. Moreover, the relationship between InPs and SPs are innovative to study. This new economic model may break the economic bottleneck of the current mobile network and bring attractive operating model.

\subsection{Ongoing research on WNV}

\begin{table*}
\newcommand{\tabincell}[2]{\begin{tabular}{@{}#1@{}}#2\end{tabular}}
\caption{Representative summary of studies on WNV}
\label{tab:WNV}
\begin{tabular}{r@{}lr@{}lr@{}lr@{}lr@{}l} \hline
\multicolumn{2}{c}{} &&year &&network type && research topic&& summary\\ \hline
\multicolumn{2}{c}{Smith \emph{et al.}\cite{NV_80211_Commodity}} && 2007 && 802.11   && Architecture  && TDM baseds wireless virtualization strategies\\
\multicolumn{2}{c}{SplitAP\cite{SplitAP}}                         && 2010 && 802.11   && Architecture  && \tabincell{l}{An virtualized architecture based on the extension\\of the virtual access point functionality in the\\physical AP}\\
\multicolumn{2}{c}{Virtual Wifi\cite{Virtual_Wifi}}               && 2011 && 802.11   && Architecture  && \tabincell{l}{achieve the goal by combining the approaches of\\both software (virtual machine and augmented\\wireless driver) and hardware (NIC)}\\
\multicolumn{2}{c}{MPAP\cite{MPAP}}                               && 2010 && 802.11   &&  Architecture && \tabincell{l}{virtualizes multiple heterogeneous wireless standards\\by adopting software radio technique and wide-band\\RF front-end hardware}\\
\multicolumn{2}{c}{Aljabari \emph{et al.}\cite{NV_WLAN_infrastructures}} && 2011 && 802.11   &&  Architecture && \tabincell{l}{A wireless networks virtualiztion approach through\\open source virtualization techniques}\\
\multicolumn{2}{c}{Matos \emph{et al.}\cite{NV_context_mesh_network}}    && 2012 && mesh     &&  Architecture && \tabincell{l}{Virtualization approach in wireless mesh network}\\
\multicolumn{2}{c}{NVS\cite{NVS_WiMAX_virtualization}\cite{NVS_Extension}} && 2010 && Wimax   &&  Architecture && \tabincell{l}{Presents an effective slice scheduler approach by\\enabling bandwidth-based and resource-based\\reservations to coexist simultaneously}\\
\multicolumn{2}{c}{Virtual basestation\cite{NV_Virtual_BS}}       && 2010 && cellular  &&  Architecture && \tabincell{l}{An wimax base station architecture based on layer-2\\switched data path and an arbitrated control path}\\
\multicolumn{2}{c}{Zaki \emph{et al.}\cite{LTE_mobile_network_virtualization}\cite{LTE_virtualization_and_pectrum_management}} && 2011 && LTE  &&  Architecture && \tabincell{l}{Air interface virtualization in LTE network by\\directly using the principle of PC virtualization}\\
\multicolumn{2}{c}{Hoffmann and Staufer\cite{NV_General_Architecture}} && 2011 && Access  &&  Architecture && \tabincell{l}{One general architecture by introducing three\\major building blocks: virtualized physical\\resources, virtual resource manager, and virtual\\network controller.}\\
\multicolumn{2}{c}{Fu and Kozat\cite{Embedding_Auction_Game}} && 2010 && Virtualization  &&  Resource allocation && \tabincell{l}{Model the resource allocation between InPs and\\SPs as a stochastic game}\\
\multicolumn{2}{c}{Park \emph{et al.}\cite{framework_virtual_embedding}} && 2009 && Virtualization  &&  Resource allocation && \tabincell{l}{A framework for resource allocation problem in\\wireless virtualization}\\
\multicolumn{2}{c}{Yun and Yi\cite{embedding_wireless_multihop_networks}} && 2011 && Virtualization  &&  Resource allocation && \tabincell{l}{Embedding problem in the wireless mesh networks}\\
\multicolumn{2}{c}{Yang \emph{et al.}\cite{Karnaugh_map}} && 2012 && Virtualization  &&  Resource allocation && \tabincell{l}{An online embedding scheme based on the dynamic\\arrival and departure scenario}\\
\multicolumn{2}{c}{Yang \emph{et al.}\cite{We_Opportunistic}} && 2013 && Virtualization  &&  Resource allocation && \tabincell{l}{An opportunistic spectrum sharing based resource\\allocation scheme for wireless virtualization}\\
\multicolumn{2}{c}{Banchs \emph{et al.}\cite{NV_Fairness_Control}} && 2012 && Virtualization  &&  Resource allocation && \tabincell{l}{An algorithm based on control theory to achieve\\the goal of both ensuring fairness in the resource\\allocation and maximizing the total throughput}\\
\multicolumn{2}{c}{GENI\cite{GENI}} && --- && Platform  &&  Platform && \tabincell{l}{Global Environment for Network Innovations}\\
\multicolumn{2}{c}{ORBIT\cite{ORBIT}} && --- && Platform  &&  Platform && \tabincell{l}{Open-Access Research Testbed for Next-Generation\\Wireless Networks}\\
\hline
\end{tabular}
\end{table*}

In its infancy, many researchers or organizations mainly focus on the virtualization of 802.11 because it is convenient for the extension from wired network virtualization. GENI envision the virtualization of wireless substrate, an extension vision of its wired platform, by deploying a unified worldwide wired-wireless virtualized testbed to enable multiple experiments sharing a common physical network\cite{GENI_Wireless}. Smith \emph{et al.}\cite{NV_80211_Commodity} propose a time division multiplexing approach wireless virtualization strategies, and address the major challenge: synchronization. Aiming at addressing the issue that a group of clients efficiently sharing the uplink spectrum blocks, SplitAP\cite{SplitAP} proposes an virtualized architecture based on the extension of the virtual access point functionality in the physical AP. Xia \emph{et al.}\cite{Virtual_Wifi} study a WLAN virtualization approach called virtual WiFi. They achieve the goal by combining the approaches of both software and hardware. Virtual WiFi supports fully functional wireless functions inside the virtual machine, and the device-specific virtual wireless functions are provided by the augmented wireless driver and the wireless NIC. MPAP\cite{MPAP} virtualizes multiple heterogeneous wireless standards by adopting software radio and wide-band RF front-end hardware. Aljabari \emph{et al.}\cite{NV_WLAN_infrastructures} introduce a wireless networks virtualization approach through open source virtualization techniques. Matos \emph{et al.}\cite{NV_context_mesh_network} studies mesh network virtualization.

Since WLAN is only a small part of mobile and wireless network, it is essential to study WNV in other wireless environments, especially cellular. NVS\cite{NVS_WiMAX_virtualization}\cite{NVS_Extension} is proposed as a network virtualization substrate for effective virtualization of wireless resources in WiMax networks. NVS presents an effective slice scheduler approach by enabling bandwidth-based and resource-based reservations to coexist simultaneously. Bhanage \emph{et al.}\cite{NV_Virtual_BS} propose \emph{virtual basestation}, which is an WiMax basestation architecture for virtualizing and integrating the ``4G'' cellular wireless network. The frame architecture is based on layer-2 switched data path and an arbitrated control path. Zaki \emph{et al.} \cite{LTE_mobile_network_virtualization}\cite{LTE_virtualization_and_pectrum_management} study the air interface virtualization in LTE network by directly using the principle of PC virtualization such as hypervisor and the virtual instances. Hoffmann and Staufer\cite{NV_General_Architecture} propose one general architecture for network virtualization in the future mobile networks by introducing three major building blocks: virtualized physical resources, virtual resource manager, and virtual network controller.

There are several studies focusing on some other fundamental problems in WNV such as virtual resource allocation and fairness controlling. Fu and Kozat\cite{Embedding_Auction_Game} model the resource allocation between InPs and SPs as a stochastic game, and find the Nash equilibrium in the conjectural prices. Given the conjectural prices, SPs have to truthfully reveal
their own value function. Park \emph{et al.}\cite{framework_virtual_embedding} propose a framework for resource allocation problem in wireless virtualization without detailed algorithm proposed. Yun and Yi\cite{embedding_wireless_multihop_networks} study the embedding problem in the wireless mesh networks. We propose a Karnaugh-map like algorithm and opportunistic spectrum sharing resource allocation solutions in \cite{Karnaugh_map} and \cite{We_Opportunistic} respectively. Banchs \emph{et al.}\cite{NV_Fairness_Control} propose an algorithm based on control theory to achieve the goal of both ensuring fairness in the resource allocation and maximizing the total throughput. And Belbekkouche \emph{et al.}\cite{Resource_allocation_survey} summarize the resource discovery and allocation problem and solutions in both wired and wireless network virtualization.

Moreover, there are several network testbeds based on WNV such as GENI and ORBIT\cite{ORBIT}. Besides, the ever-increasing mobile virtual network operators (MVNO) accelerate the development of WNV. A representative summary of WNV studies is shown in Table \ref{tab:WNV}.

\subsection{Challenges and open issues}
\subsubsection{Architecture and node virtualization}
To efficiently support future mobile and wireless network, WNV should be compatible to multiple heterogeneous networks and guarantee the isolation property. Considering that different wireless protocols require quite different implementation and performance parameters, relying on one device to implement all the functions is impracticable. Therefore, WNV needs to design a flexible architecture and node virtualization approach to address these challenges.
\subsubsection{Efficient mapping and management}
As WNV introduces virtual networks to carry services and applications, efficiently and precisely mapping these services into the most appropriate virtual networks is challenging. Moreover, service requirements and the network statuses are fast time-varying, which asks for effective and dynamic manage and control of the virtual networks and physical substrate.
\subsubsection{Virtual resource allocation}
Virtual resource allocation problem is one fundamental issue for WNV. Not only because it reflects the relationship among multiple InPs and SPs, but virtualization is also a promising way for resource sharing. However, allocation virtual resources is quite difficult because of various factors such as efficiency, fairness, selfish, and ever-varying features.
\subsubsection{Economic model}
WNV introduces an innovative economic model. Consequently, all the entities in the model undoubtedly pursue the maximal revenue and minimal cost. The complicated network condition as well as the complex relationship among these entities affect their goals. Therefore, it is important and challenging to study the economic model and further find optimal operating models.

\section{Cooperation between SDWN and WNV} \label{sec:Combination}
\begin{figure}
\centering
  \includegraphics[width=0.50\textwidth]{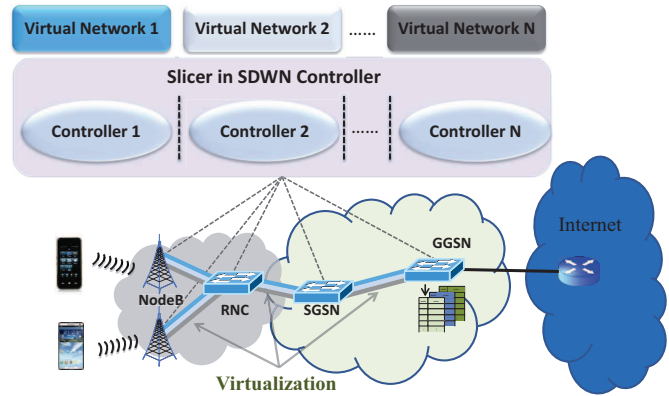}
\caption{Joint design for SDWN and WNV}
\label{fig:Combination}       
\end{figure}
\subsection{Necessity and importance of joint design}
As mentioned above, SDWN and WNV provide appropriate opportunities to integrate the ``twins'' in information technology: mobile communications and Internet. Although SDWN and WNV are considered as different technologies in motivations, goals, technical details, implement approaches, and etc., they highly complement each other and achieve a promising direction for the study on future mobile and wireless network. Concretely, the decoupling of control plane and data plane as well as introducing the centralized control plane, as proposed by SWDN, may significantly benefit the optimal control and operation strategy, facilitate the implementation, improve the programmability and customization for WNV. Meanwhile, virtualization enhances the scalability and flexibility, and improves the resource utilization for SDWN.

SDNW and WNV are mutually beneficial but not dependent on each other. Networks can be virtualized without software-defining and vice-versa. Even though, the great potential caused by the joint design between SWDN and WNV offers valuable and attractive motivations for researchers. For example, a flow based software-defined networking testbed, e.g. OpenFlow, is convenient and necessary to deploy various experiments with various requirements by implementing network virtualization.

Fig. \ref{fig:Combination} depicts one feasible joint design between SDWN and WNV. In this architecture, control plane and data plane are separated and a logically centralized control plane is introduced, which is in accordance with SDWN. Meanwhile, WNV is also supported so that it enables several concurrent virtual networks. Network slicing is integrated in the centralized control plane by providing multiple SDNW controllers, each of which corresponds to one virtual network and schedules rules of its virtual network devices. Therefore, one network device may maintain several flow tables belong to different virtual networks.

\subsection{Ongoing and potential approach}
FlowVisor\cite{FlowVisior_1}\cite{FlowVisior_2} is an important approach for the combination between software-defined networking and network virtualization. Different from common network slicing technique such as MPLS and VLAN, FlowVisor proposes a clean abstraction that virtualizes the network into multiple slices. FlowVisor is a hardware abstraction layer locating at logically between underlying forwarding hardware and the control software by utilizing OpenFlow protocol. By partitioning the flow space, e.g. flow-table, in each switch according to the corresponding flow-entries for slices, FlowVisor hosts multiple guest controllers, one controller per slice. FlowVisor guarantees the isolation feature among slices, including both the data plane isolation and control plane isolation of the slice. Consequently, one guest controller can just control and manage its own slice. Philip \emph{et al.}\cite{NV_Openflow_eNodeB} study an architecture for e-NodeB virtualization based on OpenFlow by utilizing FlowVisor.

Network Function Virtualization (NFV)\cite{NFV}\cite{NFV_SDN}, firstly proposed in the \emph{SDN and OpenFlow World Congress}, addresses several intractable problems caused by the proprietary hardware-based appliances. NFV leverages standard IT virtualisation technology to consolidate many network equipment types onto industry standard high volume servers, switches and storage, which could be located in Datacenters, Network Nodes and in the end user premises. NFV has been supposed as one potential approach to combining the software-defined networking and network virtualization.

OpenRAN\cite{OpenRadio} proposes a cloud computing based software-defined radio access network architecture. There are four types of virtualization in the architecture scheduled by the SDN controller: application level virtualization, cloud level virtualization, spectrum level virtualization, and cooperation level virtualization.

\subsection{Challenges and open issues}
\subsubsection{NFV and middlebox optimizing}
Recently, NFV enabling middlebox has captured increasing attention. Plenty of network functions may be implemented in middlebox running on the commodity hardware by virtualization. Meanwhile, middlebox may be effectively managed and controlled though SDWN. However, several key issues need to be addressed, including middlebox placement problem, flow path optimizing, dynamic resource allocation, and etc.
\subsubsection{Software-defined multi-dimension virtualization}
There are multiple dimensions of virtualization in WNV, such as spectrum resources, access devices, and forwarding devices. Various dimensions of virtualization call for quite different implementation approaches. Achieving one unified software-defined virtualization control covering all the dimensions is obviously challenging.
\subsubsection{Mobile and wireless environment}
Mobile and wireless environment is complicated and varies continuously, which makes it quite difficult to implement just one technology of SDWN and WNV, let alone the combination of both of them.
\subsubsection{Trade-off between fine-grained and implementation difficulties}
Fine-grained sofware-defining and virtualization may achieve higher performance. On the other hand, this also extremely causes implementation difficulties and further makes the vendors and operators hesitate to undertake the design and implementation. Therefore, a reasonable trade-off between fine-grained and implementation difficulties should be found.

\section{Conclusion} \label{sec:Conclusion}
Mobile and wireless network is undergoing several fundamental changes: mobile Internet has become one irreversible trend, while mobile network changes to multimedia-oriented. However, the current mobile and wireless network can hardly catch up with it due to multiple intractable challenges, which are deeply rooted in the inherent design on the mobile communication and Internet. Consequently, simply making some modifications can hardly solve these issues. On the other hand, a completed clean-slate architecture, not compatible with the current networks, is also impractical because whether operators or vendors will not take risks until they clearly understand the revenue and cost, which is nearly impossible. Therefore, future mobile and wireless network needs a revolutionary architecture while supporting the smoothly evolution.

This paper presents a survey focusing on two latest and promising technologies: software-defined wireless network (SDWN) and wireless network virtualization (WNV). SDWN and WNV significantly benefit the convergence of heterogeneous wireless networks, improve the resource utilization, facilitate the network innovations from the network layer down to physical layer, provide customized services and guarantee the QoS and QoE, and increase the revenue of all the network entities. Meanwhile, SDWN and WNV are naturally compatible with the current networks and efficiently support the smoothly evolving. Finally, implementing and combining SDWN and WNV still have lots of open issues. This requires us to solve a series of challenges step by step for the future mobile and wireless network.



\end{document}